\documentclass[11pt,a4paper]{article}
\usepackage[T1]{fontenc}
\usepackage[utf8]{inputenc}
\usepackage{authblk}

\usepackage{geometry}
\usepackage{amsfonts,amsmath,amssymb,amsthm,nicefrac}
\usepackage{ulem}
\usepackage{cancel}
\usepackage{yhmath}
\usepackage{arcs}
\usepackage{graphicx}
\usepackage{url}
\usepackage{subcaption}
\usepackage{url}
\usepackage{caption}
\usepackage{subcaption}

\title{\bf On the hierarchy of classicality and symmetry of quantum states}
\author[1,2,3]{Arsen Khvedelidze}
\author[3,4]{Astghik Torosyan}

\affil[1]{A. Razmadze Mathematical Institute, Iv. Javakhishvili Tbilisi State University, Tbilisi, Georgia}
\affil[2]{Institute of Quantum Physics and Engineering Technologies, Georgian Technical University, Tbilisi, Georgia}
\affil[3]{Laboratory of Information Technologies, Joint Institute for Nuclear Research, Dubna, Russia} 
\affil[4]{A.I. Alikhanyan National Science Laboratory (YerPhI), Yerevan, Armenia}

\date{ }

\begin{document}

\maketitle

\begin{abstract}
The interrelation between classicality/quantumness and symmetry of states is discussed within the phase-space formulation of finite-dimensional quantum systems.  We derive representations  for classicality measures $\mathcal{Q}_N[H_{\varrho}]$ of states from the stratum of given symmetry type $[H_{\varrho}]$ for the Hilbert-Schmidt ensemble of qudits. The expressions for measures are given in terms of the permanents of matrices constructed from the vertices of the special Wigner function's positivity polytope. The supposition about the  partial order  of classicality indicators $\mathcal{Q}_N[H_{\varrho}]$ in accordance with the symmetry type of stratum is formulated.
\end{abstract} 

\tableofcontents

\newpage

\section{Introduction} 

Not all things admit to be ordered, but some do. 
It is remarkable that sometimes after their ordering is recognized, the other things,  at first glance independent from the former,  reveal the corresponding order as well, thereby showing their hidden interrelations with one another. In the present note we would like to draw attention to a similar situation  occurring in statistical description of  finite-dimensional quantum systems.  Namely, we argue that  if quantum states are ordered  with respect to their ``symmetry'',  then they exhibit also  the ordering with respect to their ``classicality'' in a way that can be formulated as:
\\
\centerline{
\textit{``The larger symmetry quantum states possess, the more classical they are!''}
}
Below, attempting to alter the above sonorous utterance into the rigorous statement, we briefly recapitulate two issues --- \textbf{the equivalence and  partial order from unitary symmetry} and \textbf{classicality of states}:
\begin{itemize}
\item \textbf{Equivalence and partial order relations from the  unitary symmetry} --- the equivalence relation between
quantum states of an $N\--$level system related to the unitary group $SU(N) $ transformation. This equivalence  results in the partition of a state space  into a strata with the symmetry characterized by the partially ordered isotropy subgroups $H_\alpha \subseteq SU(N)$;

\item \textbf{Classicality of states} 
---
the notion of classical states based on the non-negativity of their quasiprobability distributions and the idea of geometric indicators of classicality $\mathcal{Q}_N[H_\alpha]$ of quantum states  defined  as the geometric probability to find a classical state on a stratum with symmetry type  $[H_\alpha]$,
\begin{equation} 
\label{eq:GlobalIndicatorStrata}
\mathcal{Q}_N[H_\alpha]= \frac{\mbox{Volume~of~classical~states~on~stratum~type~}[H_\alpha]  }
{\mbox{Volume~of~all~states~on~stratum~type~}
[H_\alpha]
}\,.
\end{equation} 
\end{itemize}
Bearing in mind the above underlying features of  partial ordering  of isotropy groups and  the corresponding classification of strata in $\mathfrak{P}_N$, we pose the question about the order of  the classicality measures (\ref{eq:GlobalIndicatorStrata}). 
Based on our computations of $\mathcal{Q}_N[H_\alpha]$ for 
3- and 4-dimensional systems we formulate the following conjecture.
\paragraph{The  hierarchy conjecture:}\textit{
Let us arrange the isotropy groups $H_\alpha $
in ascending order, starting from the maximal torus $T_N$ up to the whole group $SU(N)$,
\begin{equation}
    T_N=H_{\min} < H_1 < \cdots < H_{\max}=SU(N)\,,
\end{equation}
then the set of classicality indicators $\mathcal{Q}_N[H_\alpha]$ inherits the hierarchy,  
\begin{equation}
\label{eq:IndicHierarchy}
  \mathcal{Q}_N[T_N] < \mathcal{Q}_N[H_1] < \cdots < \mathcal{Q}_N[SU(N)]=1\,.
\end{equation}
}
In the present note we describe two methods of analytical calculations  of  measures (\ref{eq:GlobalIndicatorStrata}) for an arbitrary $N\--$level quantum system.
For the readers convenience, 
before describing these technical tools,
in the next section  we start  with  the generic issues of the unitary symmetry representation in closed quantum systems putting an accent on geometrical features of phase space description of finite-dimensional quantum systems mainly following our recent  publications 
\cite{AKh2017-WF,AKhT2019}.

\section{Symmetry and geometry}

Here we briefly  summarise  how the unitary symmetry of the underling  Hilbert space $\mathbb{C}^N$ of $N\--$dimensional quantum system
\footnote{For brevity, we will henceforth call N-level system  ``$N\--$qudit'', or simply ``qudit'',  if a specific dimension is irrelevant.}
imposes certain constraints  on geometric and statistical properties of its state space (for generic concepts see review  
\cite{MichelZhilinskii2001} and references therein).

\paragraph{The unitary symmetry, equivalence classes and partial order}
The state space $\mathfrak{P}_N$ of an $N\--$qudit  can be identified with the subspace of $N\times N $ Hermitian, trace-one  positive semidefinite matrices: 
\begin{equation}
\label{eq:StateSpace}
\mathfrak{P}_N =\{\, \varrho \in M_N(\mathbb{C}) \ |\ \varrho=\varrho^\dagger\,,\quad  \varrho \geq 0\,,  \quad \mbox{tr}\left( \varrho \right) = 1  \, \}\,.
\end{equation}
The  $U(N)$ automorphism of $\mathbb{C}^N$ induces the adjoint $SU(N)$ transformations of density matrices $\varrho \in \mathfrak{P}_N$: \begin{equation}
\label{eq:UP}
    \varrho \mapsto   \varrho^\prime = \mathrm{Ad}_g \varrho \,, \qquad  g\in SU(N)
\end{equation}
and sets up  the equivalence between points of the orbit  $ \mathcal{O}_\varrho =\{
\mathrm{Ad}_g\varrho,\ g  \in SU(N)
\}$ through the state $\varrho \in \mathfrak{P}_N.$ 
In a view of this equivalence the orbits provide partition of $\mathfrak{P}_N$, but being not locally finite (every non-empty open set intersects infinitely many orbits)  it can not serve as decomposition of $\mathfrak{P}_N$. However, with this equivalence relation there is another  kind  of partition 
named the ``orbit type'', which  is based on 
the notion of the \textit{isotropy group (stabilizer)} 
$H_x \subset SU(N)$
of point $x\in \mathfrak{P}_N$,
\[
H_x =\{\, g\in SU(N)\ | \ \mathrm{Ad}_g x = x\, \}\,.
\]
Two points $x,y \in \mathfrak{P}_N $ are declared  to be of the same type if their stabilizers are conjugate subgroups of $SU(N)$.
If the stabilizer $H_x$ of some/any point $x$ in the orbit belongs to the conjugacy class of subgroup $H$ in $SU(N)\,,$
we say that \textit{the type of the orbit} is $[H]$
and by $\mathfrak{P}_{[H_\alpha]}$ denote the set  of points of $\mathfrak{P}_N\,,$  whose stabilizer  is conjugated to the subgroup $H_\alpha$:
\begin{equation}
\mathfrak{P}_{[H_\alpha]}: =\big\{\, x \in \ \mathfrak{P}_N|\   H_x \mbox{~is~conjugate~to}\  H_\alpha \, \big\}\,.
\end{equation}
Here  $\alpha$ is the set enumerating the conjugacy classes of the isotropy groups. 
The isotropy group  of  density matrix is determined by the algebraic degeneracy of its spectrum and therefore the number of conjugacy classes is equal to the number $P(N)$ of different representations of integer $N$ as the sum of positive natural numbers, $\alpha =\{1, 2, \dots,  p(N)\}$. The subsets $\mathfrak{P}_{[\mathrm{H}_\alpha]}$ are  termed as \textit{strata} and can be partially ordered in accordance with the partial order  of the corresponding isotropy groups \footnote{If $H$ and $K$ are isotropy subgroups of $G$\,, we define a partial ordering on equivalence classes by writing $(H) < (K)$ if $H$ is $G$-conjugate to a subgroup of $K$\,. This defines a partial ordering on the set of the  isotropy types of orbits.}.
Hence we arrive at the orbit type decomposition of state space:
\begin{equation}
\label{eq:OrbitDec}
\mathfrak{P}_N=\bigcup_{\alpha}{\mathfrak{P}}_{[H_\alpha]}\,.
\end{equation}
Each stratum in 
(\ref{eq:OrbitDec})
can be described in terms of states with a fixed degeneracy as follows.
Consider $(N-1)\--$dimensional simplex $C_{N-1}$ of ordered eigenvalues:
\begin{equation}
\label{eq:NorderedSim}
 C_{N-1} := \{\, \boldsymbol{r} \in \mathbb{R}^N \, \biggl| \, 
\sum_{i=1}^{N} r_i = 1\,, \quad 1\geq r_1\geq r_2 \geq \dots \geq r_{N-1}\geq r_N \geq 0 \, \}\,. 
\end{equation}
For our further aims it is enough to restrict ourselves by considering  only the  positive density matrices of maximal rank, i.e. remove from  the simplex the subset  $\{r_1=1\} \cup  \{r_N=0\}\,.$  
This truncated simplex 
is a union of eigenvalues of non-singular density matrices  of the fixed degeneracy $\boldsymbol{k}=(k_1, k_2, \dots, k_n ),$ 
\begin{equation}
\label{eq:DegSet}
\mathfrak{P}_{\boldsymbol{k}} = \{\, \varrho\in\mathfrak{P}_N\,, k_i \in \mathbb{Z}_+\, |\, 
\det(\varrho-\lambda)=\prod_{i=1}^n (r_i-\lambda)^{k_i}\,, \quad   \sum_{i=1}^n k_i= N \, \}\,.
\end{equation}
Finally, taking into account an admissible transposition of eigenvalues,  we arrive at the decomposition of a given stratum:
\begin{equation}
\label{eq:stratumDeg}
\mathfrak{P}_{[H_\alpha]}=
 \bigcup_{\omega \in S_n }\mathfrak{P}_{\omega.\boldsymbol{k}}\,.   
 \end{equation}
In (\ref{eq:stratumDeg})
by $\omega\cdot\boldsymbol{k}=
\{ k_{\omega(1)},
 k_{\omega(2)},
 \dots, k_{\omega(n)}\}
$ we denote the action of a symmetric group $S_n$ on a given partition of $N$ into $n$ natural numbers  $k_1, k_2, \dots, k_n\,.$ 
\paragraph{Unitary invariance of  probability distributions on strata}
Let us assume that the probability density function of the qudit ensemble  is invariant under (\ref{eq:UP}):
\begin{equation}
\label{eq:InvPDF}
    P(\varrho)= P(g\varrho g^\dagger)\,, \qquad  \forall\ g \in SU(N)\,.
\end{equation}
Due to the  invariance property (\ref{eq:InvPDF}) one can get convinced that the probability density function 
on a given stratum $\mathfrak{P}_{[H_\alpha]}$
reduces to the following 
expressions:
\begin{equation}
\label{eq:UEstrat}
P(\varrho)=
\sum_{\omega \in S_n }
P_{\omega\cdot\boldsymbol{k}}(r_1,\dots,r_n)\,
\, \mathrm{d}r_1\wedge \cdots \wedge \mathrm{d}r_N\wedge \mathrm{d}\mu_{U(N)/H}\,,
\end{equation}
which shows that the measure factorizes into the factor  corresponding to the  measure on
subset $\mathfrak{P}_{\boldsymbol{k}} $ of the simplex $\mathcal{C}_{N-1}$ and   the Haar  measure on the coset 
$U(N)/H$\,. 

\paragraph{The Hilbert-Schmidt ensemble of qudits on principal stratum } One of the widely used unitary invariant probability  density function originates from the Hilbert-Schmidt (HS) metric  on $\mathfrak{P}_N:$
\begin{equation}
 \label{eq:HSGen}
\mathrm{g_{{}_\mathrm{HS}}} \propto \mathrm{Tr} \left(\mathrm{d}\varrho\otimes\mathrm{d}\varrho\right)\,.
\end{equation}
If a density matrix $\varrho$ belongs to the principal stratum with maximal torus isotropy group, $\varrho \in\mathfrak{P}_{[T^{(N-1)}]}$, then the  metric (\ref{eq:HSGen}) defines the standard \textit{Hilbert-Schmidt ensemble} of random full-rank $N\--$qudits with the well-known joint probability distribution of distinct eigenvalues,
\begin{equation}
P(r_1,\dots,r_N) \propto \,
    \delta(1-\sum_{j=1}^N r_j)  \prod_{j<k}^N (r_j-r_k)^2\,.
\end{equation}

\paragraph{The Hilbert-Schmidt ensemble of qudits on degenerate strata} If the full-rank density matrix is degenerate with multiplicity of eigenvalues 
$\{k_1, k_2, \dots,  k_n\}$, i.e., its isotropy group is $H=U(k_1)\times\cdots\times U(k_n)\,,$ then 
the  joint probability distribution of eigenvalues reads: 
\begin{equation}
\label{eq:HSstrat}
P_{k_1, \dots, k_s}(r_1, \dots, r_s) \propto 
 \delta(1- \sum_{i=1}^{n} k_i r_i)    \prod_{i<j}^{1\dots n} (r_i-r_j)^{2 k_i k_j}\,.
 \end{equation}

\section{Classicality and geometry} 

In this section we formulate the notion of classicality of qudit states as an  existence of a corresponding proper probability distributions. Namely,  we relate the  classicality with the Wigner function (WF) positivity and describe the underlying  geometry of the state space. 
In our consideration  we use  the $(N-2)$-parametric family WFs given by the dual pairing of a  density matrix $\varrho$ and Stratonovich-Weyl (SW) matrix valued kernel $\Delta(\boldsymbol{z} | \boldsymbol{\nu})$  on  the phase space 
$\Omega_N$ (cf. for details in \cite{AKh2017-WF,AKhT2019}):
\begin{equation}
W^{(\boldsymbol{\nu})}_\varrho(\boldsymbol{z}) = \mbox{tr}
    (\varrho
\Delta(\boldsymbol{z}  
    |
\boldsymbol{\nu}))\,,
\qquad\boldsymbol{z} \in \Omega_N\,.
\end{equation}

\paragraph{Classical states and WF positivity polytope} The ``classical states'' form the subset $\mathfrak{P}_N^{\mathrm{Cl}} \subset  \mathfrak{P}_N$ of states  whose Wigner function 
$W^{(\boldsymbol{\nu})}_\varrho(z) $ in a given representation with moduli parameters 
$\boldsymbol{\nu}=\{\nu_1, \nu_2, \dots, \nu_{N-2}\}$
is non-negative everywhere over the phase space: 
\begin{equation}
\label{eq:P+}
\mathfrak{P}^{\mathrm{Cl}}_N = \{\, \varrho \in \mathfrak{P}_N\,\ | \  W^{(\boldsymbol{\nu})}_\varrho(z) \geq 0\,, \quad  \forall z\in \Omega_N \, \}.
\end{equation}
The ``classical states on a fixed stratum'' $ \mathfrak{P}_{H_\alpha}$ are defined respectively  as:
\begin{equation}
\label{eq:P+Stratum}
\mathfrak{P}^{\mathrm{Cl}}_{H_\alpha}=
\mathfrak{P}^{\mathrm{Cl}}_N
\cap
\mathfrak{P}_{H_\alpha}\,.
\end{equation}
In order to describe explicitly the classical states (\ref{eq:P+}) and  (\ref{eq:P+Stratum}) one can consider  the following  linear functional $\mathfrak{P}_N \to \mathbb{R}$: 
\begin{equation}
\label{eq:supF}
    w[\varrho] := \inf_{g \in U(N)} \, W^{(\boldsymbol{\nu})}_{g \varrho g^\dagger}\, (z)\,
\end{equation}
and exploit the following
observation.

\noindent{\bf Proposition I.}
\textit{
The  zero-level set  of functional 
$w[\varrho]$,
\begin{equation}
    H_N:\  \{ \varrho \in \mathfrak{P}_N\,|\,
     w[\varrho] =0 \}\,, 
\end{equation}
describes  the supporting hyperplane 
\begin{equation}
\label{eq:SuppHyper}
(\boldsymbol{r}^\downarrow, \boldsymbol{\pi}^\uparrow) = r_1\pi_{N}+r_2\pi_{N-1} +\dots +r_{N}\pi_1 =0 \, 
\end{equation}
of the  convex set 
of classical states.
The tuples $\boldsymbol{r}^\downarrow=\{r_1, r_2, \dots, r_N\}$  and $\boldsymbol{\pi}^\downarrow=\{\pi_1, \pi_2, \dots, \pi_N\}$  in 
(\ref{eq:SuppHyper})
denote  the eigenvalues of the density matrix $\varrho$ and 
the SW kernel $\Delta(\boldsymbol{z}\,|\,\boldsymbol{\nu})$ respectively, both arranged in decreasing  order. The  SW kernel eigenvalues 
$\boldsymbol{\pi}$ satisfy the following equations: 
\begin{equation}
\label{eq:specdelta}
    \sum_{i}^{N}\, \pi_i =1\,, \qquad 
\sum_{i}^{N}\, \pi_i^2 =N\,.
\end{equation}
}

\noindent{\bf Proposition II.}
\textit{
The intersection of the hyperplane $H $ with 
the  simplex 
(\ref{eq:NorderedSim})
defines the Wigner function's positivity polytope corresponding to the canonical projection $p: 
\mathfrak{P}_{N} \mapsto \mathfrak{P}_N/ SU(N)
$  of the classical states.}

The Propositions I and II follow from the results 
of \cite{AKhT2019},  where 
the image of classical subsets $\mathfrak{P}^{(\mathrm{Cl})}_{[H_\alpha]}$  under the canonical quotient mapping where introduced:  
\begin{eqnarray}
\label{eq:OritP+}
\mathcal{C}_{N-1}^\ast(H_\alpha)=  \{\, p(x)\, \ | \ x \in \mathfrak{P}_{H_\alpha}^{\mathrm{Cl}}\,\}\,. 
\end{eqnarray}

\paragraph{The set of classicality measures.}
A knowledge of the WF positivity  polytope allows one to extract information on the classicality/quantumness of states. 
Based on the definition of region of classical states (\ref{eq:P+}), we can define sequence of classicality/quantumness indicators evaluating relative weight of the classical states. Namely one can consider the collection of different  geometric probabilities  of finding a classical state in a given  unitary invariant statistical ensemble (\ref{eq:UEstrat}), among them \cite{AKhT2020,Abbaslietal2020}: 
\begin{enumerate}
\item The \textit{global indicator of classicality of ensemble},
\begin{equation}
\label{eq:QDef}
\mathcal{Q}_N=
\frac{\int_{\mathfrak{P}_N^{\mathrm{Cl}}}\,\mathrm{d}\mu}{\int_{\mathfrak{P}_N}\mathrm{d}\mu}\,.
\end{equation}
\item The \textit{indicator of classicality  of a stratum ensemble},
\begin{equation}
\label{eq:QDefStratun}
\mathcal{Q}_N[H_\alpha]=
\frac{\int_{ \mathfrak{P}^{\mathrm{Cl}}_{[H_\alpha]}}\,
    \mathrm{d}\mu}{\int_{\mathfrak{P}_{[H_\alpha]}}\,\mathrm{d}\mu}\,.
\end{equation}
\end{enumerate}
Here it is in order to make a few comments,  in  (\ref{eq:QDef})-(\ref{eq:QDefStratun})
the measure $\mathrm{d}\mu $ is assumed to be the unitary invariant  of the form 
(\ref{eq:UEstrat}). In the subsequent section we will specify  the measure corresponding to the ensemble of Hilbert-Schmidt  qudits (\ref{eq:HSGen}) and (\ref{eq:HSstrat})
 for $N=2,3,4$, i.e. qubit, qutrit, and quatrit respectively. 
 Note that  we expect that
 $\mathcal{Q}_N=
 \mathcal{Q}_N[H_0]$,
 since the principal stratum
 with $H_0= U(1)^N$ 
 differs from the whole space state 
 $\mathfrak{P}_N$ by 
 a measure-zero set only.

\section{Computing the indicators}

According to (\ref{eq:stratumDeg}), the stratum $\mathfrak{P}_{[H_\alpha]}$ consists from subsets of matrices with a certain degeneracy type. Due to the unitary invariance of probability distribution functions (\ref{eq:InvPDF}), any above introduced classicality indicator depends only on the joint  probability distribution of eigenvalues of the density matrix and thus can be rewritten as:
\begin{equation}
\label{eq:QOrb}
\mathcal{Q}_N [H_\alpha]=
\frac{\sum_{\omega \in S_s }\int_{\mathcal{C}^{\ast}_{N-1}(H_\alpha)}
P_{k_\omega(1),\dots,k_\omega(s)}(r_1,\dots,r_s)\,
\mathrm{d}r_1\wedge\dots\wedge\mathrm{d}r_s}{\sum_{\omega \in S_s }\
\int_{\mathcal{C}_{N-1}(H_\alpha)}
 P_{k_\omega(1),\dots,k_\omega(s)}(r_1,\dots,r_s)\,
\mathrm{d}r_1\wedge\dots\wedge\mathrm{d}r_s}\,.  
\end{equation}
In (\ref{eq:QOrb}) the integral in the denominator represents the volume of the orbit space of stratum $\mathfrak{P}_{[H_\alpha]}$\,. The integration in the nominator of (\ref{eq:QOrb}) is over the 
WF positivity polytope 
$\mathcal{C}_{N-1}^\ast(H_\alpha)$:
\begin{equation}
\mathcal{C}_{N-1}^\ast(H_\alpha) =  \
\left\{\, \boldsymbol{\pi} \in \mbox{\bf spec}\left(\Delta(\Omega_N)\right) \ \, |\ \,  ( \boldsymbol{r}^\downarrow, \boldsymbol{\pi}^\uparrow) \geq 0,  \quad \forall\, \boldsymbol{r} \in \mathcal{C}_{N-1}(H_\alpha) \, \right\}\,.
\end{equation}
Hence, for the  
Hilbert-Schmidt qudits with  probability distribution functions (\ref{eq:HSGen}) and (\ref{eq:HSstrat}) the 
evaluation  of the classicality indicators reduces to the problem of integration of polynomials over the convex polytopes.

\paragraph{Simplicial decomposition} 
It is known that computation of the volume of polytopes of varying dimension is {\sf \#P}-hard and that even approximating the volume is hard \cite{DyerFrieze1988}.  
Currently, the most powerful method  for an efficient  approximation of integrals 
in (\ref{eq:QOrb}) over polytopes remains the Monte Carlo-type algorithms.
However, often when the polytopes are functions  of parameters (as in our case, when its structure depends on representation of SW kernel), an exact analytical calculations  of the volume is requested, the  situation becomes extremely complicated.
In this case the computational methods stem from the observation that a convex polytope admits decomposition  into a union of simplices, satisfying certain properties. Based on this  idea of triangulation, the polytope volume might  be computed either summing up volumes of simplices or using the signed decomposition methods if a given polytope is decomposed into signed simplices such that the signed sum of their volumes gives  the volume of the polytope. Leaving aside the question of an efficient simplicial decomposition, below we  describe two 
methods of evaluation of integrals from homogeneous polynomials over the simplex. 
\paragraph{The 1st  Lasserre-Avrachenkov (LA) method of integration over simplex}
We are interested in calculation  of the integral of  the  polynomial $p(x_1, x_2, \dots, x_n)$ over the $n$-simplex  ${C}_n \in \mathbb{R}^{n}$
with vertices 
$\boldsymbol{v}_{0},  \boldsymbol{v}_{1}, \dots, 
\boldsymbol{v}_{n}$
\begin{equation}
\label{eq:homqn}
V\left(p; \, C_n \right)  = \int_{C_n}p (\boldsymbol{x})
\mathrm{d}\boldsymbol{x}
\end{equation}
with respect  to  the $n\--$dimensional Lebesgue  measure.
With this aim we recall an elegant analytical method reducing calculation of integrals from homogeneous 
polynomials to the integration of the corresponding polarization form of those polynomials \cite{LasserreAvrachenkov2001}.
Briefly it can be stated as follows. 
Let $p(\boldsymbol{x})$ be  $q$\--homogeneous polynomial, 
\begin{equation}
p: \mathbb{R}^n \to \mathbb{R}\,, \qquad p(t\boldsymbol{x})=t^qp(\boldsymbol{x})\,, \quad \forall \, t\in \mathbb{R}~\mbox{and}~ \boldsymbol{x} \in \mathbb{R}^n\,, 
\end{equation}
and let  $H_p(\boldsymbol{X}_1, \boldsymbol{X}_2, \dots , \boldsymbol{X}_q)$ be the polarization of $p$, the mapping   
$
(\mathbb{R}^n)^q \mapsto \mathbb{R}
\,, $
which is symmetric q-linear form 
such that \footnote{
The well-known formula, 
\begin{equation}
H_p(\boldsymbol{X}_1, \boldsymbol{X}_2, \dots , \boldsymbol{X}_q)=
\frac{1}{q!}\frac{\partial}{\partial t_1}\frac{\partial}{\partial t_2}\cdots \frac{\partial}{\partial t_q}p(t_1\boldsymbol{X}_1 +t_2\boldsymbol{X}_2 +\dots +t_q\boldsymbol{X}_q)\biggl|_{\boldsymbol{t}=0}\,,
\end{equation}
gives a compact representation for the polarization.}
\begin{equation}
H_p(\boldsymbol{x}, \boldsymbol{x}, \dots , \boldsymbol{x})=p
(\boldsymbol{x}).
\end{equation}
According to the Lasserre-Avrachenkov theorem \cite{LasserreAvrachenkov2001},  the integration in 
(\ref{eq:homqn})
results in  summation of the values of polarization $H_p(\boldsymbol{X}_1, \boldsymbol{X}_2, \dots , \boldsymbol{X}_q)$
evaluated at  the vertices 
$\boldsymbol{v}_{0},  \boldsymbol{v}_{1}, \dots, 
\boldsymbol{v}_{n}$
of a simplex:
\begin{equation}
\label{eq:LAintegral}
    {V}\left(p;\, {{C}_n}\right) =\frac{\mathrm{vol}({C}_n)}{\binom{n+q}{n}
    }\,\sum_{\sum
    _{0}^{n}a_i =q}
{H}_p(
\overbrace{\boldsymbol{v}_{0},\dots, \boldsymbol{v}_{0}}^{a_0}, 
\overbrace{\boldsymbol{v}_{1},\dots, \boldsymbol{v}_{1}}^{a_1},\dots, \overbrace{\boldsymbol{v}_{n},
\dots,\boldsymbol{v}_{n}}^{a_n})\,.
\end{equation}

\paragraph{The 2nd Lasserre  method of integration over simplex}
Another important for us result has been noted by J.Lassere 
\cite{Lasserre2021}. He 
proved that  integrating a polynomial of degree $q$ on an arbitrary simplex (with respect to Lebesgue measure) reduces to evaluating $q$\--homogeneous polynomials of degree $j=1,2, \dots , q$
each at a unique point $\boldsymbol{s}_j$ of the simplex. 
Bearing in mind that the integration over an arbitrary simplex can be reduced to the integration over the 
canonical simplex \footnote{
The canonical $n\--$simplex $K_n
\subset \mathrm{R}^n$ is defined as  
\(
    K_n=\{ \boldsymbol{x} \in 
    \mathrm{R}^n_+ \,|\, x_1+x_2+\dots +x_n \leq 1
    \}.
\)
}
by  a certain affine transformation,  we give the 
formulation of the method for the canonical simplex case. 
Namely, 
let the polynomial $p(\boldsymbol{x})$ of degree $q$ be rewritten as $p(\boldsymbol{x})=\sum_{j=0}^q\,p_j
(\boldsymbol{x})\,,$ where 
$p_j(\boldsymbol{x})=\sum_{|\alpha|=j} p_\alpha \boldsymbol{x}^\alpha$ is a homogeneous polynomial of degree $j$.  Then according to \cite{Lasserre2021} the 
integration over the canonical $n\--$dimensional simplex $K_n$
gives
\begin{equation}
\label{eq:Lassere}
 \int_{K_n} p(\boldsymbol{y}) \mathrm{d}
 \boldsymbol{y}= \mathrm{vol}(K)\left(
 \hat{p}_0 +
\sum_{j=1}^q\,\hat{p}_j(\boldsymbol{s}_j)\right)\,, 
\end{equation}
where $\boldsymbol{s}_j=\frac{1}{\sqrt[j]{(n+1)\dots (n+j)}}(1,1,\dots, 1)$
and $\hat{p}(\boldsymbol{x})$
 stands for the associated “Bombieri” polynomial:
\begin{equation}
    \hat{p}(\boldsymbol{x})= 
    \sum_{\alpha \in \mathbb{N}^n} 
    p_\alpha \alpha_1 !\dots \alpha_n !\, \boldsymbol{x}^\alpha\,, \qquad \alpha=(\alpha_1, \alpha_2, \dots , \alpha_n)\,.
\end{equation}
Note that expression  (\ref{eq:Lassere}) 
differs from the well-known cubature formulae.
In (\ref{eq:Lassere})
instead of evaluating a single polynomial  at several points, as it takes place in the case of cubature formulae, one evaluates  polynomials of degree $j$ at a single point only.

\paragraph{Applying methods to the Hilbert-Schmidt measure} Both the above mentioned methods of integration can be used analyzing the classicality indicators $\mathcal{Q}_N[H_\alpha]$  of the Hilbert-Schmidt ensembles of qudits. 
Here we outline the general scheme of calculation while in the next section considering low-dimensional systems $N=2,3,4,$ some principal technical details will be elucidated.
As a first step, we decompose the WF positivity polytope into the sum of simplices, i.e., 
\begin{equation}
    Q_N[T_{N}]=
\sum_{\mathrm{simplices}}\, I_C(\boldsymbol{\pi})\,,
\end{equation}
where the typical element of the  sum  is integral over a certan $n$\--simplex ${C}_{n}$
given as the convex hull of $n$ vertices $\mathcal{C}_n(\boldsymbol{\pi}): = \mathrm{conv}(v_0, v_1(\boldsymbol{\pi}), \dots, v_{n}(\boldsymbol{\pi}))$:
\begin{equation}
\label{eq:tint}
I_C(\boldsymbol{\pi}) \propto
\int_{C_n(\boldsymbol{\pi})}
\mathrm{d}\boldsymbol{r} \, \delta(1-\sum_i^{n+1} r_i)
\prod_{i < j}^{n+1}\,(r_i-r_j)^2\,.
\end{equation}
Note that in the case we are interested in, the vertices $v_i(\boldsymbol{\pi})$ are rational functions of the SW kernel
eigenvalues. Their  exact form  follows from the separating hyperplane equation 
(\ref{eq:SuppHyper}).
The integrand in (\ref{eq:tint}) due to $\delta\--$function factor is not a homogeneous polynomial and thus the LA 
formula (\ref{eq:LAintegral})
is not applicable directly.
But, using the map from the canonical (standard) simplex $K_{n}$ 
to the simplex 
${C}_{n}:$  
\begin{equation}
\label{eq:mapKC}
 K_{n} \mapsto {C}_{n}(\boldsymbol{\pi}):\,\ \boldsymbol{r}= \boldsymbol{v}_0+ \sum_{\alpha=1}^{n}
(\boldsymbol{v}_\alpha(\boldsymbol{\pi})-\boldsymbol{v}_0)u_\alpha\,,    \end{equation}
the integral  reduces to the integral over the canonical 
$n\--$simplex 
\begin{equation}
\label{eq:tint2}
I_C(\boldsymbol{\pi}) \propto
\mathrm{vol_E}(C_{n}(\boldsymbol{\pi}))\,
\int_{K_{n}}
\mathrm{d}\boldsymbol{u}\, \mu(\boldsymbol{u})\,,
\end{equation}
where $\mathrm{vol_E}({C}_{n}(\boldsymbol{\pi}))$ denotes the Euclidean volume of the simplex ${C}_{n}(\boldsymbol{\pi})$
\footnote{The Euclidean volume of 
$n\--$simplex in $\mathbb{R}^n$ in terms of $(n+1)$\--vertices  reads:
\[
\mathrm{vol_E}({C}_{n}):= \frac{1}{n!}\biggl|  \det
\begin{pmatrix}
 \boldsymbol{v}_0 &\boldsymbol{v}_1 &\dots  &\boldsymbol{v}_n \\ 
 1&1&\dots &1
\end{pmatrix}
\biggl|
\,.
 \]
 }
 and 
$ \mu(\boldsymbol{u})$ is homogeneous polynomial of order $q=n(n-1)$:
\begin{eqnarray}
 \label{eq:hompol}
\mu(\boldsymbol{u})=\prod_{i<j}\left(\sum_{\alpha=1}^{n}(
v_\alpha^i(\boldsymbol{\pi})-v_\alpha^j(\boldsymbol{\pi})\,u_\alpha)
 \right)^2\,.
\end{eqnarray}
The polynomial (\ref{eq:hompol})
can be rewritten  as 
\begin{equation}
\label{eq:HSqform}
\mu(\boldsymbol{u})= (-1)^n\,\prod_{s=1}^q \,(\boldsymbol{\alpha}_s, \boldsymbol{l}(\boldsymbol{u}))
\,,
\end{equation}
 where 
\begin{equation}
\boldsymbol{l}(\boldsymbol{u})^i=\sum_{\alpha=1}^{n} v_\alpha^i u_\alpha\,, 
\end{equation} 
and  $\boldsymbol{\alpha}_s = \{\boldsymbol{e}_i-\boldsymbol{e}_j\,, \
i,j=  1,2, \dots n
\} $ are
$n\--$vectors  constructed out of the standard unit $n$\--dimensional vectors $\boldsymbol{e}_i$.
Linearity of 
$\boldsymbol{l}(\boldsymbol{u})^i$ implies 
q-linearity of  the associated to the polynomial $\mu(\boldsymbol{u})$ polarization form  $H$, 
\begin{equation}
\label{eq:polformHS}
{H}_\mu(\boldsymbol{X}_1, \boldsymbol{X}_2, \dots , \boldsymbol{X}_{q}):= \frac{1}{q!}\sum_{\sigma \in S_q}\,
\prod_{s=1}^q\, (\boldsymbol{\alpha}_s,\,\boldsymbol{l}(\boldsymbol{X}_{\sigma(s)}))\,.
\end{equation}
The expression (\ref{eq:polformHS}) shows that the polarization form corresponding to the Hilbert-Schmidt measure is given by the normalized  permanent of $q\times q$ matrix,  
\begin{equation}
 H_p(\boldsymbol{X}_1\, \boldsymbol{X}_2, \dots , \boldsymbol{X}_{q})= \frac{1}{q!}\,\mathrm{perm}||(\boldsymbol{\alpha}_i,\,\boldsymbol{l}(\boldsymbol{X}_j)||\,. 
\end{equation}
Hence, using the LA formula  
(\ref{eq:LAintegral})
and noting that  
$\boldsymbol{l}(\boldsymbol{e}_j)=\boldsymbol{v}_j$,
we arrive at 
\begin{equation}
\label{eq:tintff}
I_C(\boldsymbol{\pi}) \propto
\frac{\kappa(\boldsymbol{\pi})}{q!}\,
\sum_{\sum_{i=1}^{n}a_i =q}
\mathrm{perm}\left||\mathrm{M}(a_1, a_2, \dots, a_{n})\right||\,,
\end{equation}
where   
 \begin{equation}
  \kappa(\boldsymbol{\pi})=   \frac{\mathrm{vol_E}({C}_{n}(\boldsymbol{\pi}))\mathrm{vol_E}(K_{n})}{\binom{n+q}{q}}\,,
 \end{equation}
and
$q\times q $ matrices $\mathrm{M}:$ 
\begin{equation}
\mathrm{M}_{st}(a_1, a_2, \dots, a_{n}):=(\boldsymbol{\alpha}_s,\,
\boldsymbol{V}_t)\,, \quad s,t=1,2,\dots, q\,,
\end{equation}
constructed out of tuples
$V=\{
\overbrace{\boldsymbol{v}_{0},\dots, \boldsymbol{v}_{0}}^{a_0}, 
\overbrace{\boldsymbol{v}_{1},\dots, \boldsymbol{v}_{1}}^{a_1},\dots, \overbrace{\boldsymbol{v}_{{n}},
\dots,\boldsymbol{v}_{n}}^{a_{n}}
\}$ for
all admissible partitions  of degree  of homogeneity $\sum a_i =q$ in integers. 

We finalize this paragraph  
noting that  the above scheme of calculations is applicable to the evaluation of the classicality indicators  $\mathcal{Q}[H_\alpha]$ for the lower-dimensional strata as well.

\subsection{Qubit}

The ordered eigenvalue simplex of a qubit represents the line segment  in $\mathbb{R}^2$:
\[C_1:\, \quad  \{r_1+r_2=1, 1\geq r_1\geq r_2\geq 0\}\,.
\] 
This interval is convex hull of  points $\boldsymbol{v}_0=\{1/2, 1/2 \}$ and  $\boldsymbol{v}_1=\{1, 0\}\,.$ 
Among qubit states the maximally mixed state at vertex $\boldsymbol{v}_0$ has maximal symmetry, the $SU(2)$ isotropy group, while  all the other states $\varrho \in \mathfrak{P}_2\,$ have the torus $\mathrm{T}^2 \in SU(2)$ as their isotropy group.
Noting that for the maximally mixed state $\varrho_\ast = 1/2 \, \mathbb{I}_2$ the Wigner function is positive, we can formally assign the  value  one to 
the classicality indicator,  $\mathcal{Q}_2[SU(2)]=1\,.$
The indicator $\mathcal{Q}_{[\mathrm{T}^2]}$ for the principle stratum can be 
calculated along the methods described in previous section  noting that the spectrum of  SW kernel is uniquely determined from 
(\ref{eq:specdelta}): 
\begin{equation}
\label{eq:qubitSW}
\pi_1=\frac{1+\sqrt{3}}{2}\,, \qquad \pi_2=\frac{1-\sqrt{3}}{2}\,,
\end{equation}
and the  supporting hyperplane 
$\pi_1r_2+\pi_2r_1
=0$ intersects the segment $C_1$ at $
\boldsymbol{v}_1
(\boldsymbol{\pi})
=
\{ \frac{1}{2}+\frac{1}{2\sqrt{3}},  
\frac{1}{2}-\frac{1}{2\sqrt{3}}\}\,.$ The integration over the intervals 
is trivial and as a result the qubit global indicator of classicality  is 
\begin{equation}
\label{eq:Qqubit}
\mathcal{Q}_{[\mathrm{T}^2]}=\frac{1}{3\sqrt{3}}\,.
\end{equation}

\subsection{Qutrit}

\paragraph{Unitary strata of qutrit state space}

The ordered eigenvalue simplex of qutrit is triangle   in $\mathbb{R}^3:$
\[C_2:\, \quad  \{r_1+r_2+r_3=1,\qquad  1\geq r_1\geq r_2\geq  r_3 \geq 0\}\,.
\] 
It is convex hull of three  points $\boldsymbol{v}_0=\{1/3, 1/3, 1/3 \}$\,, $\boldsymbol{v}_1=\{1/2, 1/2, 0\}$  
and $\boldsymbol{v}_2=\{1, 0, 0\}$\,.
The possible multiplicity of eigenvalues are 
$\boldsymbol{k}=(1,1,1)$\,, $\boldsymbol{k}=(1,2)$ and $\boldsymbol{k}=(2,1)$\,, and there are three corresponding strata of $\mathfrak{P}_3$:
\begin{itemize}
\item the 8-dimensional principal stratum with isotropy class $[\mathrm {T}^3]$ consisting   of matrices with a simple spectrum,  $1> r_1\neq r_2\neq r_3 > 0$,
\[
\mathfrak{P}_{[\mathrm {T}^3]}: \quad
\{
\varrho \in \mathfrak{P}_{3} \, |\,  \mbox{spec}(\varrho):=
(r_1, r_2, r_3 )\,, 1 > r_1 > r_2 >r_3 > 0
\};
\]
\item the 5-dimensional degenerate stratum with isotropy class $[\mathrm {S(U(2)\times U(1))}]$ is the locus of density matrices with the degeneracies $\boldsymbol{k}=(2,1)$ and $\boldsymbol{k}=(1,2)$\,,
\[
\mathfrak{P}_{[\mathrm {S(U(2)\times U(1))}]}: \quad
\mathfrak{P}_{1,2} \bigcup \mathfrak{P}_{2,1}\,,
\]
with components 
\begin{eqnarray}
\mathfrak{P}_{1,2}: &&
\{ 
\varrho \in \mathfrak{P}_{3} \, |\,  \mbox{spec}(\varrho):=
(r_1, r_2, r_3 )\,,1>r_1\neq r_2=r_3> 0
\},  \nonumber \\
 \mathfrak{P}_{2,1}: &&
\{ 
\varrho \in \mathfrak{P}_{3} \, |\,  \mbox{spec}(\varrho):=
(r_1, r_2, r_3 )\,,  1>r_1=r_2\neq r_3>0
\};
\nonumber
\end{eqnarray}
\item the 0-dimensional stratum, $\mathfrak{P}_{[SU(3)]}$, the  mixed state with the triple degeneracy  $\boldsymbol{k}=(3)$\,, $r_1=r_2=r_3=1/3$\,.
\end{itemize}
\begin{figure}[h!]
\center{
\includegraphics[width=0.5\textwidth]{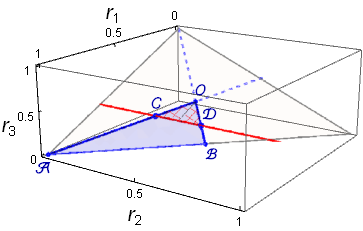} 
}
\caption{Triangle $\triangle AOB$ as the ordered 2-simplex of qutrit eigenvalues, and the hatched triangle $\triangle COD$ corresponds to the classical states. 
}
\label{fig:2Simplex}
\end{figure} 

\paragraph{Global indicator $\mathcal{Q}_{[\mathrm{T}^3]}$ } 
The regular stratum $\mathfrak{P}_{[T^3]}$ consists of  density matrices with a simple spectrum: $
     1 > r_1  > r_2 > r_3\geq 0\,, \quad 
    \sum_{i=1}^3 r_i = 1.
 $
The plane separating  classical and quantum states of qutrits, 
\begin{equation}
H_3: \quad 
\pi_1r_3+\pi_2r_2+\pi_3r_1=0
\end{equation}
intersects the partially ordered  simplex of qutrit eigenvalues by the straight line passing through  the points 
\begin{equation}
\label{eq:vertecCD}
C=\frac{1}{3\pi_3-1}(\pi_3-1,\pi_3, \pi_3)\,, \qquad  
D=\frac{1}{3\pi_1-1}(\pi_1, \pi_1, \pi_1-1)\,.
\end{equation}
Hence, the eigenvalues of qutrit classical states belong to WF positivity triangle $\triangle COD$ with  the   vertices (\ref{eq:vertecCD}) 
and the vertex of maximally mixed 
state $\boldsymbol{O}=(\frac{1}{3}, \frac{1}{3}, \frac{1}{3})$  with triple 
degeneracy. Note that
\begin{eqnarray}
&& \min_{\boldsymbol{\pi}}|OD|=\frac{1}{2\sqrt{6}}\ 
    \mbox{at}\ \boldsymbol{\pi}=\frac{1}{3}(5,-1,-1),
   \quad 
   \max_{\boldsymbol{\pi}}|OD|=\frac{1}{\sqrt{6}}
   \ \mbox{at}\  \boldsymbol{\pi}=(1,1,-1),
   \nonumber
\\
 && \min_{\boldsymbol{\pi}}|OC|=\frac{1}{2\sqrt{6}}
 \ \mbox{at}\  \boldsymbol{\pi}=\frac{1}{3}(5,-1,-1),
 \quad  \max_{\boldsymbol{\pi}}|OC|=\frac{1}{2\sqrt{2}}
\ \mbox{at}\  \boldsymbol{\pi}=(1,1,-1),
\nonumber
\end{eqnarray}
and  the line $H_3$ is tangent to the disc of the  ``absolutely'' classical states: $ r \leq \frac{1}{4}\,.$
Here we use the relation 
$r_1^2+r_2^2+r_3^2=1/3+2/3r^2\,$ 
between eigenvalues of a qutrit and its   Bloch radius $r$.

Following the suggested generic scheme, the evaluation  of volume of classical states of qutrit reduces to the integration over 
the WF positivity triangle $\triangle COD$. The integrand of equivalent canonical $K_2\--$simplex is given by a sextic homogeneous polynomial of the following form:
\begin{eqnarray}
\label{eq:int3}
  I(\boldsymbol{\pi})&\propto& \frac{1}{(3\pi_1-1)^3(1-3\pi_3)^3}\,\int_{K_2}du
 dv\, u^2 v^2\left(
 \frac{u}{3\pi_1-1}+\frac{v}{1-3\pi_3}
 \right)^2 \,.
 \end{eqnarray}
Based on the 2nd Lasserre method (\ref{eq:Lassere}), we
evaluate the associated Bombieri polynomial at point $\boldsymbol{s}_6=(\frac{2}{8!})^{1/6}\,(1,1)$  and arrive at the following exact expression for the indicator $\mathcal{Q}_3={I(\boldsymbol{\pi})}/{I(1, 0, 0)}$:
 \begin{equation}
\mathcal{Q}_{[T^3]}=\frac{1}{(3\pi_1-1)^3(1-3\pi_3)^3}\left[
\frac{4}{(3\pi_1-1)^2}+\frac{4}{(1-3\pi_3)^2}+\frac{6}{(3\pi_1-1)(1-3\pi_3)}\right]
 \end{equation}
 for all possible SW kernels of qutrit states from the principle stratum.

\paragraph{$\mathcal{Q}_3\--$indicator of qutrits from degenerate stratum.} 
The stratum 
$\mathfrak{P}_{[S(U(2)\times U(1))]}$ has two pieces, associated to density matrices with the degenerate  eigenvalues $r_1=r_2\neq r_3$ and $r_1\neq r_2= r_3$\,, respectively.
Hence,  the $\mathcal{Q}_3\--$indicator for the degenerate stratum of a qutrit reads:
\begin{equation}
\label{eq:Q3U2U1}
\mathcal{Q}_{[S(U(2)\times U(1))]}=\frac{\mathrm{vol_{HS}}(\mathfrak{P}^{\mathrm{Cl}}_{1,2} \bigcup \mathfrak{P}^{\mathrm{Cl}}_{2,1})}{\mathrm{vol_{HS}}\left(\mathfrak{P}_{1,2} \bigcup \mathfrak{P}_{2,1}\right)}\,.  
\end{equation}
Exploiting the suggested techniques of  integration for (\ref{eq:Q3U2U1}), we obtain:
\begin{equation}
\label{eq:QD}
   \mathcal{Q}_{[S(U(2)\times U(1))]} = \frac{2^5}{1+2^5} \left(\frac{1}{(3\pi_1-1)^5} + \frac{1}{(1-3\pi_3)^5}\right)\,.
\end{equation}

\begin{figure}[h!]
\center{
\includegraphics[width=0.4\textwidth]{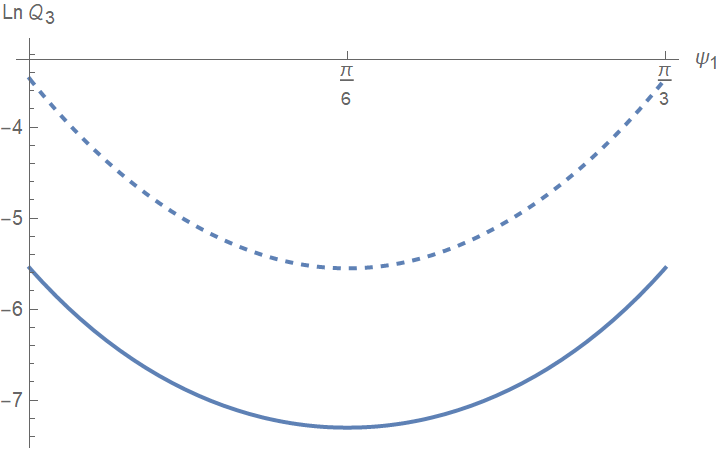} 
}
\caption[]{Comparing indicators of qutrit regular  (solid curve) and degenerate strata (dashed curve).}
\label{fig:Qutrit-Q3-Ln}
\end{figure}

\paragraph{Order relations between indicators } Now we are in position to compare the classicality indicators for different strata.
Introducing the $\zeta\--$angle parameterization ($\zeta\in [0,\pi/3]$) 
for the SW kernel eigenvalues 
\begin{equation} 
\label{eq:N3PI}
\pi_1=\frac{1}{3}+\frac{2}{\sqrt{3}}\sin\zeta+
\frac{2}{3}\cos\zeta\,,\
\pi_2=\frac{1}{3}-\frac{2}{\sqrt{3}}\sin\zeta+
\frac{2}{3}\cos\zeta\,,\
\pi_3=\frac{1}{3}1-\frac{4}{3}\cos\zeta\,, 
\end{equation}
one can easily verify the inequalities for the classicality indicators of qutrit: 
\begin{equation}
 0 <   \mathcal{Q}_{[\mathrm{T}^3]}   <  \mathcal{Q}_{[S(U(2)\times U(1))]} < 1
\,.
\end{equation}
The Fig. 
\ref{fig:Qutrit-Q3-Ln}  demonstrates how 
the partial order of the corresponding  isotropy groups
$[\mathrm{T}^3] < [S(U(2)\times U(1))]$
is reproduced at the level of their classicality indicators.

\subsection{Quatrit}


\paragraph{$\mathcal{Q}_4\--$indicator for 
quatrit regular stratum}
The orbit  space of quatrit represents tetrahedron and the stratum $\mathfrak{P}_{[\mathrm{T}^4]}$
is given by the density matrices with the regular spectrum,
$
1 > r_1 > r_2 > r_3 > r_4 \geq 0 \,, \ 
    \sum_{i=1}^4 r_i = 1\,. 
$
In order to describe the subset of classical states of quatrit,  we analyse intersections of 3-simplex with the supporting plane 
\begin{equation}
\label{eq:plane}
    H_4: \qquad \{\boldsymbol{r} \in \mathcal{C}_3\,, \boldsymbol{\pi} \in \mathcal{P}_3 \, |\, \pi_1=(\pi_1-\pi_4) r_1 + (\pi_1-\pi_3) r_2 + (\pi_1-\pi_2) r_3\}\,,
\end{equation}
where 
\begin{equation}
 \mathcal{P}_3 : \qquad \sum_{i=1}^3\,\pi_i  =1\,, \quad 
 \sum_{i=1}^3\,\pi_i^2  =4\,.
\end{equation}
The possible cross-sections of the  plane (\ref{eq:plane}) with
the tetrahedron are either a triangular, or a quadrilateral depending on the moduli  space $\mathcal{P}_3$. Indeed, one can see that the maximally mixed state $\boldsymbol{r}_{*}=(1/4, 1/4, 1/4, 1/4 ) $ has positive WF and the rays emanating from $\boldsymbol{r}_{*}$ along the edges of tetrahedron intersect the plane $H_4$ at three points. Then there are two possibilities and hence, only two 
types of admissible cross-sections:
\begin{enumerate}
\item[\bf (A)] triangles, if the intersection points belong to edges of the tetrahedron  emanating from vertex of maximally mixed states;
\item[\bf (B)] quadrilaterals, if an intersection point lies outside the edge of the tetrahedron.
\end{enumerate}
An explicit form of intersection points,  taking into
account the eigenvalues order $\pi_1\geq \pi_2 \geq \pi_3 \geq \pi_4\,,  $ are: 
\begin{enumerate}
    \item Intersection with edge $OC$ at point with symmetry $ S(U(3)\times U(1))$: 
\begin{equation}
    \boldsymbol{P}_{{}_{OC}}=\frac{1}{4\pi_1-1}
    \begin{pmatrix}
       \pi_1 \\
       \pi_1 \\
       \pi_1 \\
       \pi_1-1
    \end{pmatrix}\,, \quad \mbox{if} \quad  \pi_1\geq 1;
\end{equation}
\item There is no intersection with the edge $AB$.
{The plane $H_4$ intersects the ray passing through the edge $AB$:
\begin{equation}
\boldsymbol{P}_{{}_{AB}}=\frac{1}{\pi_3-\pi_4}
    \begin{pmatrix}
       \pi_3 \\
       -\pi_4 \\
       0 \\
       0
    \end{pmatrix};
\end{equation}
and this point belongs to the edge $AB$ if  $$\pi_3 > \pi_4\, \& \, \pi_4 <0\, \& \,   
            \pi_3+\pi_4 > 0\, \&  \, \pi_3 > 0 \,,$$ but 
these  conditions never hold}.

\item
Intersection with the edge $AC$ at point 
with symmetry $U(1)\times SU(2) \times U(1)$:
\begin{equation}
\boldsymbol{P}_{{}_{AC}}=
\frac{1}{1-\pi_1-3\pi_4}
    \begin{pmatrix}
       1-(\pi_1+\pi_4) \\
       -\pi_4\\
       -\pi_4 \\
       0
    \end{pmatrix}
    \,, \quad \mbox{if } \quad  
    \pi_1\leq 1\,  \&\,
    \pi_4\leq 0;
\end{equation}
\item
Intersection with edge $AO$ at point 
with symmetry $S(U(1)\times U(3))$:
\begin{equation}
    \boldsymbol{P}_{{}_{OA}}=\frac{1}{1-4\pi_4}
    \begin{pmatrix}
       1-\pi_4 \\
       -\pi_4 \\
       -\pi_4 \\
       -\pi_4
    \end{pmatrix}\,, \quad \mbox{if } \quad  \pi_4 \leq 0;
\end{equation}

\item Intersection with the edge $OB$ at point with symmetry $S(U(2)\times U(2) )$:
\begin{equation}
\boldsymbol{P}_{{}_{OB}}=
\frac{1}{2((\pi_1+\pi_2)-(\pi_3+\pi_4))}
    \begin{pmatrix}
       (\pi_1+\pi_2) \\
       (\pi_1+\pi_2)\\
       -(\pi_3+\pi_4)\\
        -(\pi_3+\pi_4)
    \end{pmatrix}\,,
    \quad \mbox{if} \quad
    \pi_3+\pi_4 \leq 0\,;
\end{equation}

\item
{
Intersection with the edge $BC$ at point 
$S(U(2)\times U(1) \times U(1) )$:
\begin{equation}
\boldsymbol{P}_{{}_{BC}}=
\frac{1}{\pi_1+3\pi_2-1}
    \begin{pmatrix}
       \pi_2 \\
       \pi_2\\
       (\pi_1+\pi_2)-1\\
       0
    \end{pmatrix}
    \,, \quad \mbox{if} \quad
    \frac{1}{2}\leq \pi_2 \leq 1\,  \&\,
    \pi_2\leq \pi_1\leq 1\,.
\end{equation}
}
\end{enumerate}

As we will see below, the {\bf A}-type configurations have either the maximal symmetry groups, 
$SU(4)$, or sub-maximal, $S(U(1)\times U(3)), S(U(3)\times 
U(1)$ and $ S(U(2)\times U(2))$ respectively, 
while for the {\bf B}-type configurations, when  the cross-section of separating plane with the simplex of quatrit eigenvalues represents a quadrilateral,  the isotropy  groups are   $S(U(1)\times U(2)\times U(1)), S(U(2)\times U(1)^2)$.

\begin{figure}[h!]
\centering
\includegraphics[width=0.5\textwidth]{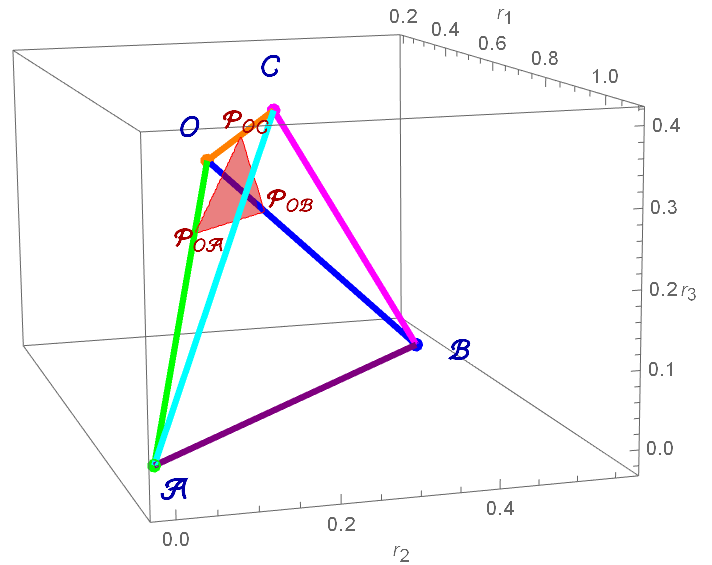} 
\caption[]{An ordered quatrit 3-simplex with the vertices,
$\boldsymbol{O}=\left(\frac{1}{4}, \frac{1}{4}, \frac{1}{4}, \frac{1}{4}\right),  
\boldsymbol{C}=\left(\frac{1}{3}, \frac{1}{3}, \frac{1}{3}, 0\right), 
\boldsymbol{B}=\left(\frac{1}{2}, \frac{1}{2}, 0, 0\right)$,  
 and $\boldsymbol{A}=\left(1, 0, 0, 0\right)\,
$ and WF positivity simplex $\mathrm{conv}\left(\boldsymbol{O, P_{OA}, P_{OB}, P_{OC}}\right)\,.$}
\label{fig:Thetrahedron1}
\end{figure}

\paragraph{WF positivity polytope  of A-type}
For this  class of SW kernels $\pi_1\geq 1$ the cross-section WF positivity polytope is a 3-simplex 
(see Fig. \ref{fig:Thetrahedron1}).
Following the suggested method,
in order to compute the H-S volume of classical states, we map the WF positivity simplex \--- the 
$\mathrm{conv}\left(\boldsymbol{O, P_{OA}, P_{OB}, P_{OC}}\right)$ \--- to a canonical 3-simplex $K_3$.
As a result, we arrive at calculation of the following integral: 
\begin{eqnarray}
\label{eq:int4-1}
I(\boldsymbol{\pi})
 &\propto&  \frac{1}{32 (4 \pi_1-1)^3 (1-4 \pi_4)^3  (\pi_1+\pi_2-\pi_3-\pi_4)^3} \, \times \\ 
\nonumber 
&& \int_{K_3}du\,dv\,dt\,u^2v^2t^2\,\left( \frac{u}{4 \pi _1-1}+\frac{t}{1-4 \pi _4}+\frac{v}{2 \left(\pi _1+\pi _2-\pi _3-\pi _4\right)} \right)^2 \, \times \\ 
\nonumber 
&& \left(\frac{u}{4 \pi _1-1}+\frac{v}{2 \left(\pi _1+\pi_2-\pi_3-\pi_4\right)} \right)^2\left( \frac{t}{1-4 \pi_4}+\frac{v}{2 \left(\pi_1+\pi _2-\pi_3-\pi_4\right)}\right)^2 \,. 
\end{eqnarray} 
Evaluating then the associated Bombieri polynomial at $\boldsymbol{s}_{12}= \left(\frac{6}{15!}\right)^{1/12}\,(1,1,1)$ and using the Lasserre formula (\ref{eq:Lassere}), we  arrive at the following expression for (\ref{eq:int4-1}): 
\begin{eqnarray}
\label{eq:k3simp}
&& 
I(\boldsymbol{\pi}) \propto \frac{1}{(4 \pi_1-1)^3 (1-4 \pi_4)^3  (\pi_1+\pi_2-\pi_3-\pi_4)^3} \times \\ 
\nonumber 
&& \Big[ 
\frac{480}{\left(4 \pi _1-1\right){}^2 \left(1-4 \pi _4\right){}^4} + 
\frac{480}{\left(4 \pi _1-1\right){}^4 \left(1-4 \pi _4\right){}^2} + 
\frac{35}{\left(\pi _1+\pi _2-\pi _3-\pi _4\right){}^6} + \\ 
\nonumber 
&& \frac{105}{\left(4 \pi _1-1\right) \left(\pi _1+\pi _2-\pi _3-\pi _4\right){}^5} + 
\frac{105}{\left(1-4 \pi _4\right) \left(\pi _1+\pi _2-\pi _3-\pi _4\right){}^5} + \\ 
\nonumber 
&& \frac{180}{\left(4 \pi _1-1\right){}^2 \left(\pi _1+\pi _2-\pi _3-\pi _4\right){}^4} + 
\frac{180}{\left(1-4 \pi _4\right){}^2 \left(\pi _1+\pi _2-\pi _3-\pi _4\right){}^4} + \\ 
\nonumber 
&& \frac{200}{\left(4 \pi _1-1\right){}^3 \left(\pi _1+\pi _2-\pi _3-\pi _4\right){}^3} + 
\frac{540}{\left(4 \pi _1-1\right) \left(1-4 \pi _4\right){}^2 \left(\pi _1+\pi _2-\pi _3-\pi _4\right){}^3} + \\ 
\nonumber 
&& \frac{540}{\left(4 \pi _1-1\right){}^2 \left(1-4 \pi _4\right) \left(\pi _1+\pi _2-\pi _3-\pi _4\right){}^3} + 
\frac{120}{\left(4 \pi _1-1\right){}^4 \left(\pi _1+\pi _2-\pi _3-\pi _4\right){}^2} + \\ 
\nonumber 
&& \frac{120}{\left(1-4 \pi _4\right){}^4 \left(\pi _1+\pi _2-\pi _3-\pi _4\right){}^2} + 
\frac{912}{\left(4 \pi _1-1\right){}^2 \left(1-4 \pi _4\right){}^2 \left(\pi _1+\pi _2-\pi _3-\pi _4\right){}^2} + \\ 
\nonumber 
&& \frac{360}{\left(4 \pi _1-1\right) \left(1-4 \pi _4\right){}^4 \left(\pi _1+\pi _2-\pi _3-\pi _4\right)} + 
 \frac{960}{\left(4 \pi _1-1\right){}^2 \left(1-4 \pi _4\right){}^3 \left(\pi _1+\pi _2-\pi _3-\pi _4\right)} + \\ 
\nonumber 
&& \frac{960}{\left(4 \pi _1-1\right){}^3 \left(1-4 \pi _4\right){}^2 \left(\pi _1+\pi _2-\pi _3-\pi _4\right)} + 
\frac{800}{\left(4 \pi _1-1\right){}^3 \left(1-4 \pi _4\right){}^3} + \\ 
\nonumber 
&& \frac{600}{\left(4 \pi _1-1\right) \left(\pi _1+\pi _2-\pi _3-\pi _4\right){}^2 \left(1-4 \pi _4\right){}^3} + 
\frac{200}{\left(\pi _1+\pi _2-\pi _3-\pi _4\right){}^3 \left(1-4 \pi _4\right){}^3} + \\ 
\nonumber 
&& \frac{315}{\left(4 \pi _1-1\right) \left(\pi _1+\pi _2-\pi _3-\pi _4\right){}^4 \left(1-4 \pi _4\right)} + 
\frac{600}{\left(4 \pi _1-1\right){}^3 \left(\pi _1+\pi _2-\pi _3-\pi _4\right){}^2 \left(1-4 \pi _4\right)} + \\ 
\nonumber 
&& \frac{360}{\left(4 \pi _1-1\right){}^4 \left(\pi _1+\pi _2-\pi _3-\pi _4\right) \left(1-4 \pi _4\right)}
\Big] \,.
\end{eqnarray}    

\begin{figure}[h!]
\centering
\begin{subfigure}[b]{0.4\textwidth}
\centering
\includegraphics[width=\textwidth]{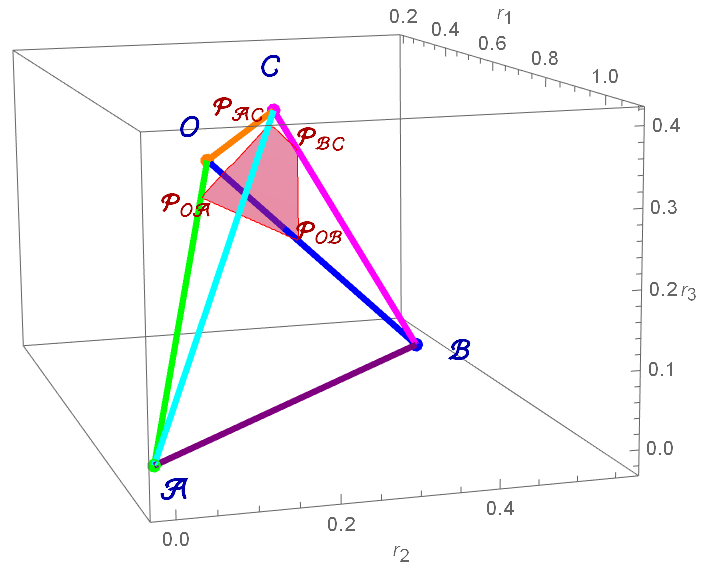}
\caption{ }
\label{fig:QuadriltrlSimplx-a}
\end{subfigure}
\begin{subfigure}[b]{0.4\textwidth}
\centering
\includegraphics[width=\textwidth]{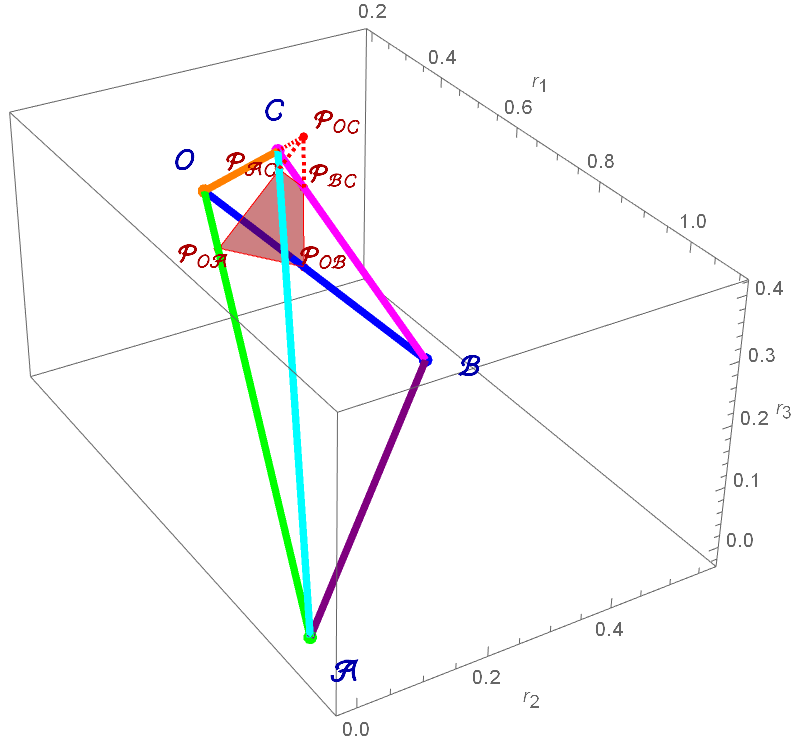}
        \caption{ }
\label{fig:QuadriltrlSimplx2-b}
\end{subfigure}
\caption[]{ The WF positivity 3-polytope formed by a cross-section ($\boldsymbol{P}_{{}_{AC}} \boldsymbol{P}_{{}_{BC}} \boldsymbol{P}_{{}_{OB}} \boldsymbol{P}_{{}_{OA}}$) of the quatrit simplex and the maximally mixed state\,.}
\label{fig:QuadriltrlSimplx}
\end{figure}

\paragraph{WF positivity polytope of B-type}
For the class of SW kernels with 
$\frac{1}{4}\leq \pi_1< 1$
the cross-section of the 
separating hyperplane of quatrit with the ordered 3-simplex of eigenvalues represents the quadrilateral which is the base of the WF positivity cone with vertex at maximally mixed state $\boldsymbol{O}$  depicted in Fig. \ref{fig:QuadriltrlSimplx}.
For computation of the H-S volume  of WF positivity polytope one can use either its decomposition into simplicies or signed simplices. An example illustrating the signed simplices decomposition is shown in 
Fig. \ref{fig:QuadriltrlSimplx2-b}, 
\begin{eqnarray}
\label{eq:int4-2}
\mbox{Vol}\big[\boldsymbol{O} \boldsymbol{C}\boldsymbol{P}_{{}_{AC}} \boldsymbol{P}_{{}_{BC}} \boldsymbol{P}_{{}_{OB}} \boldsymbol{P}_{{}_{OA}}\big] = 
\mbox{Vol}\big[\boldsymbol{O}\boldsymbol{P}_{{}_{OC}} \boldsymbol{P}_{{}_{OA}} \boldsymbol{P}_{{}_{OB}}\big] - 
\mbox{Vol}\big[\boldsymbol{C}\boldsymbol{P}_{{}_{OC}} \boldsymbol{P}_{{}_{AC}} \boldsymbol{P}_{{}_{BC}}\big]\,.
\end{eqnarray} 
Using  the LA method of computation, we obtain the representation for the classicality indicators in the form of the piecewise rational functions of the SW kernels eigenvalues.  Due to the combinatorial complexity, the corresponding expressions are too cumbersome to be written explicitly in the text. However, being interested in comparing the classicality indicators $\mathcal{Q}_4$  for different strata in relations to their symmetry type, we can effectively use these expressions.  
In the next section we briefly  summarize the relevant observations.

\section{Summary}

Our calculations reveal 
interrelation between hierarchy of  quantum states symmetry and their classicality/quantumness  which in our opinion deserve a certain attention.

\begin{figure}[h!]
\begin{minipage}[h]{0.32\linewidth}
\center{\includegraphics[width=0.25\linewidth]{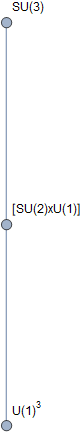}}
\end{minipage}
\hfill
\begin{minipage}[h]{0.32\linewidth}
\center{\includegraphics[width=1\linewidth]{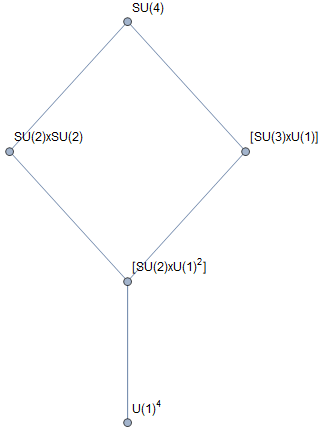}}
\end{minipage}
\hfill
\begin{minipage}[h]{0.32\linewidth}
\center{\includegraphics[width=1\linewidth]{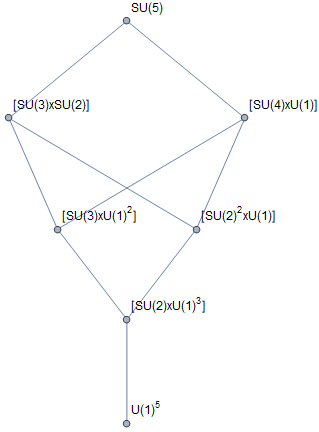}}
\end{minipage}
\begin{minipage}[h]{0.96\linewidth}
\begin{tabular}{p{0.32\linewidth}p{0.32\linewidth}p{0.32\linewidth}}
\centering 
\footnotesize $N=3$ & \centering 
\footnotesize $N=4$ & \centering 
\footnotesize $N=5$ \\
\end{tabular} 
\end{minipage}
\caption[]{Hasse diagram for SU(N) group, $N=3,4,5$.}
\label{fig:Hasse}
\end{figure}

We  found that the classicality indicators of qutrit and quatrit for the regular stratum and degenerate strata respect the order of the corresponding isotropy groups in agreement  with their  Hasse diagram  for partially ordered subgroups of unitary groups (Fig. \ref{fig:Hasse}).

\begin{figure}[h!]
\centering
\includegraphics[width=0.8\textwidth]{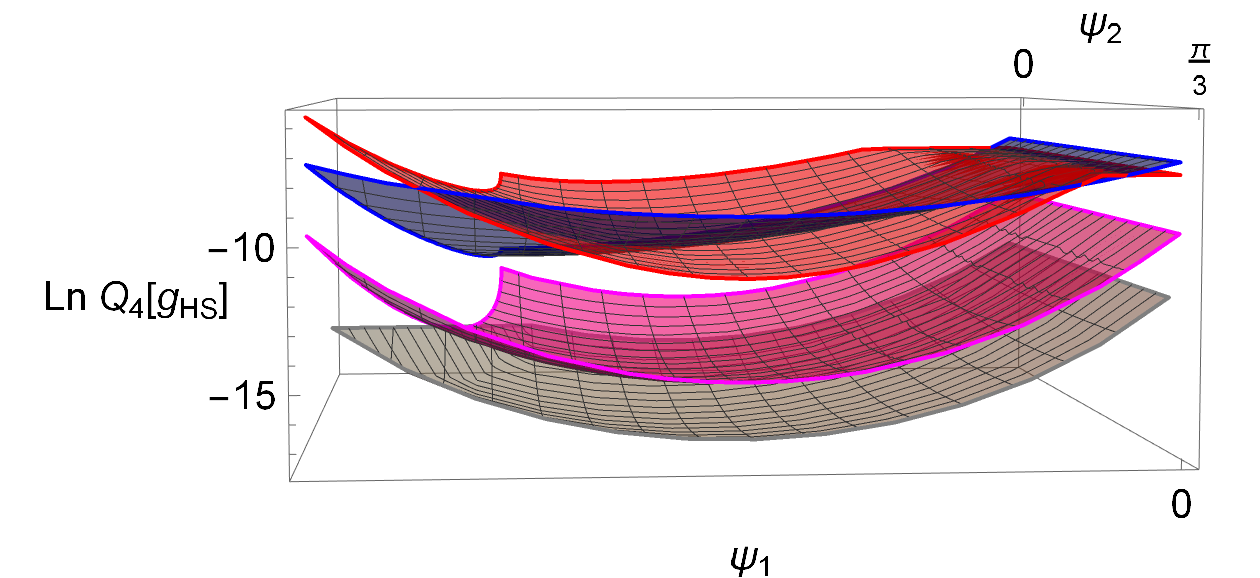}
\caption[]{$\mathcal{Q}_4\--$indicators for strata of different isotropy types:  
$\mathcal{Q}_{[S(U(3)\times U(1))]}$ (red surface); $\mathcal{Q}_{[U(1)\times SU(2)\times U(1)]}$ (blue surface); $\mathcal{Q}_{[SU(2)\times U(1)^2]}$ (magenta surface); and regular $\mathcal{Q}_{[T^4]}$-indicator (gray surface).}
\label{fig:Quatrit-Q-Ln}
\end{figure}

\begin{figure}[h!]
\begin{minipage}[h]{0.32\linewidth}
\center{\includegraphics[width=1\linewidth]{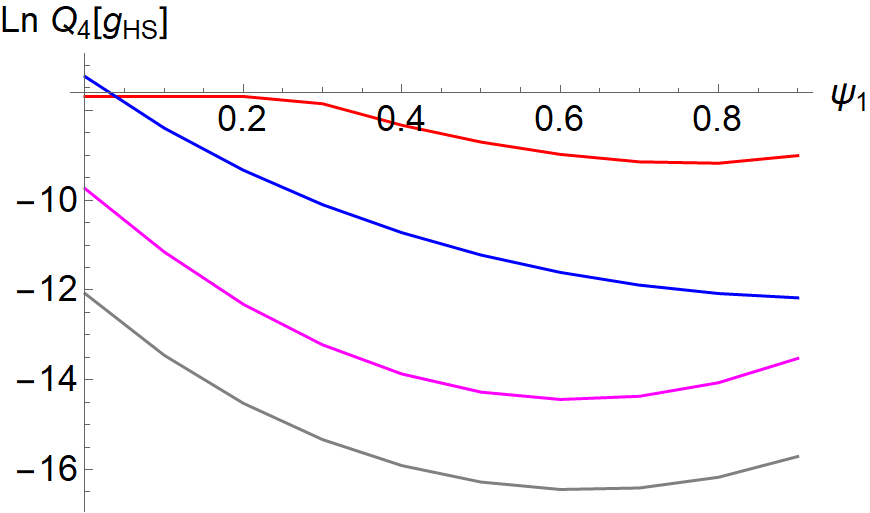}}
\end{minipage}
\hfill
\begin{minipage}[h]{0.32\linewidth}
\center{\includegraphics[width=1\linewidth]{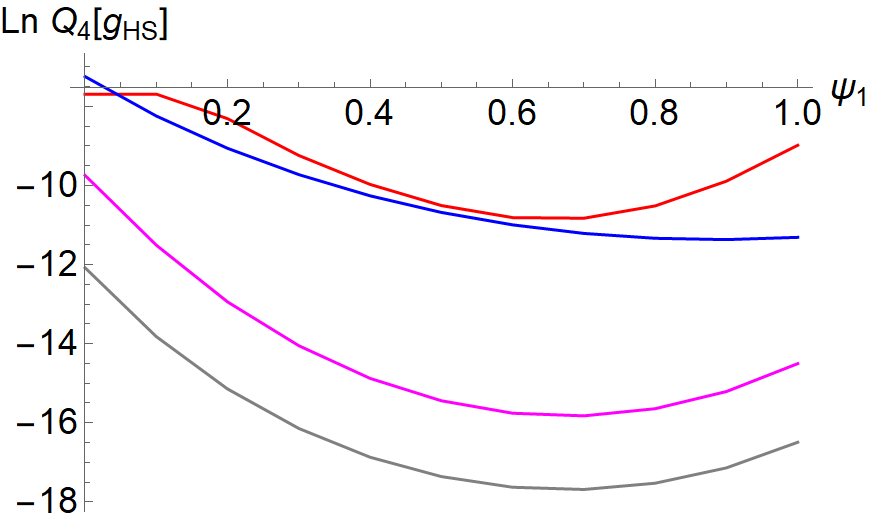}}
\end{minipage}
\hfill
\begin{minipage}[h]{0.32\linewidth}
\center{\includegraphics[width=1\linewidth]{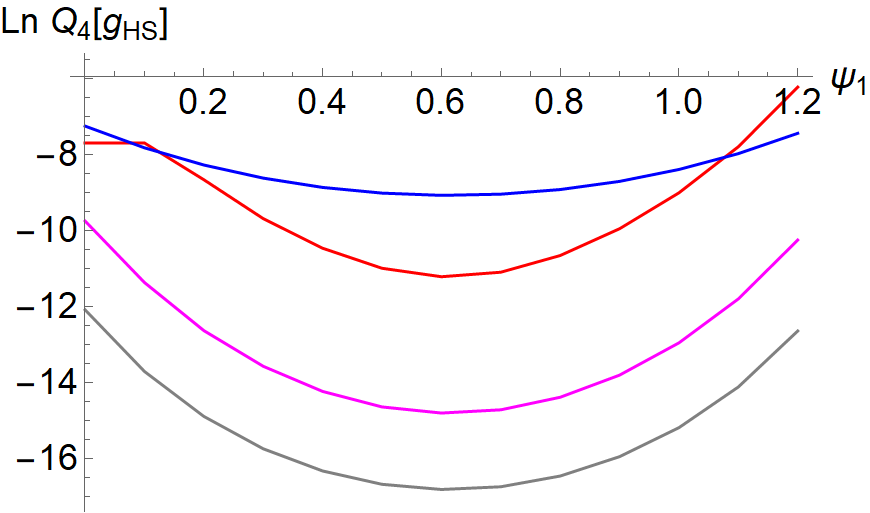}}
\end{minipage}
\begin{minipage}[h]{0.96\linewidth}
\begin{tabular}{p{0.32\linewidth}p{0.32\linewidth}p{0.32\linewidth}}
\centering 
\footnotesize $\psi_2=0$ & \centering 
\footnotesize $\psi_2=\pi/6$ & \centering 
\footnotesize $\psi_2=\pi/3$ \\
\end{tabular}
\end{minipage} 
\caption[]{A quatrit classicality indicators  for different strata as a function of WF moduli parameter $\psi_1$ for the fixed values 
$\psi_2=\{0, \frac{\pi}{6}, \frac{\pi}{3}\}$.}
\label{fig:QuatritPsi1}
\end{figure}

The curves for qutrit $\mathcal{Q}_3\-- $indicators  in Fig. \ref{fig:Qutrit-Q3-Ln} 
and the surfaces in Fig. \ref{fig:Quatrit-Q-Ln} describing   quatrit $\mathcal{Q}_4\--$indicators for all possible strata as function of a quatrit moduli parameters $\psi_1$  and $\psi_2\,$ \footnote{These quatrit moduli parameters are angles of the M{\"o}bius spherical triangle $(2,3,3)$ on a unit sphere
(cf. \cite{AKhT2019}).
} illustrate the  mentioned hierarchical structure  of classicality in relation with the symmetry properties of states.
Making the corresponding slices of $\mathcal{Q}_4\--$indicators for the fixed values of the moduli parameter $\psi_2=\{0,\pi/6, \pi/3\}$  in Fig. \ref{fig:QuatritPsi1}, 
we distinctly see that for the groups at the same ``level'' in Hasse diagram the values of
$\mathcal{Q}$\--indicators are of the same order (even equal for certain WF representations), otherwise their magnitudes significantly vary.
 
\section*{Acknowledgments}
The work of A.K. has been partially supported by the Shota Rustaveli National Science Foundation of Georgia, Grant. The research was partially supported by the Higher Education and Committee of MESCS RA (Research project № 23/2IRF-1C003).


\end{document}